\documentclass[prb,preprint]{revtex4-1}
\pdfoutput=1

\usepackage{amsmath}  
\usepackage{amsfonts} 
\usepackage{graphicx} 
\usepackage{color}
\usepackage{verbatim}
\usepackage{float}
\usepackage{multirow}
\usepackage{footnote}
\usepackage{threeparttable}
\usepackage{subfigure}

\begin{document}


\title{A Novel Approach for Exploring the Light Traveling Path in the Medium with a Spherically Symmetric Refractive Index}

\author{\textbf{Shengyang Zhuang, Jiaqi Yin and Jun Li$^*$}\\
~\\
{S. Zhuang$^1$, J. Yin$^2$, J. Li$^3$}\\
{$1$ School of Astronautics} \\ 
{$2$ School of Electronics and Information Engineering}\\ 
{$3$ School of Physics}\\ 
{Harbin Institute of Technology, Harbin 150001, China}\\ 
{Email: lijuna@hit.edu.cn}}


\begin{abstract}
A unique perspective approach based on an analogy method is presented to solve the ray equation in a model of a continuous inhomogeneous medium, which has a spherically symmetric distribution. Basically, in the standard undergraduate physics teaching, the curved ray path caused by refraction in a medium with a continuously varying refractive index has always been a relatively difficult problem to solve. The equation is usually expressed in terms of partial differential equations (PDEs), which cannot be solved by
analytical methods. Based on the analogy method, this work proposes the correspondence between ray refraction in an established medium model and the inverse-square central force system, succinctly obtaining their relation equations mathematically. We also verify the correctness of the method by qualitative and quantitative analysis. In terms of theoretical validation, we analyse the relation between Fermat's principle and Hamilton's principle, which lays a theoretical foundation for the analogy method. In addition, ray paths in the medium model were also simulated by numerical calculations based on COMSOL Multiphysics, and the results are in perfect agreement with the  conclusions. 

\end{abstract}

\maketitle 
\section{Introduction} 
Fermat's principle derives the conclusion that in homogeneous media the refractive index $n$ is constant and light rays travel along straight lines.\cite{feynman} However in graded index media in which $n$ depends on the spatial coordinates, light rays propagate along curved paths. For example, in the field of laser ranging, due to the atmospheric effects, uneven refractive index causes the laser propagation path to bend, making the apparent distance, elevation angle, and velocity of the target measured by the lidar different from its true distance, elevation angle and velocity.\cite{laser} In the field of photoelectric detection, due to the influence of atmospheric refraction, the reconnaissance beam travels along a curve instead of a straight line.\cite{photoelectric} In the field of astronomical observation, atmospheric refraction is an important factor affecting the accuracy of astronomical measurement and celestial navigation.\cite{astronomical} Mirage, a natural phenomenon, is a well-known illusion caused by the refraction of distant objects, resulting in a temperature gradient as the sun warms the ground. Consequently, light bends at the temperature gradient boundary due to different densities and refractive indices.\cite{hecht,lehn}

These problems are essentially the solution of the optical path equation in the continuous inhomogeneous media. We could presume such a medium exists, featuring a refractive index with the spherically symmetric distribution. In fact, there are no shortage of such models and spaces,\cite{lehn1985} for instance, partial regions of the refractive index of the earth atmosphere is symmetrically distributed in concentric circles\cite{atmosphere}. RL White et al.\cite{white,qian,wu} used the phenomenon of stellar light refraction through the atmosphere to navigate the satellite autonomously. Jennifer C. Ricklin et al.\cite{ricklin} studied the effects of the atmospheric channel on laser communication. Liu et al.\cite{liu} studied the influence of atmospheric refraction on the laser timing deviation in dual-wavelength space. In fact, research about light ray paths in spherically symmetric continuous inhomogeneous media has always been a significant subject to discuss. In previous methods, this problem is generally solved by building mathematical models, which are usually expressed in terms of partial differential equations (PDEs)\cite{math,springer}. In this paper, a novel method of analogy between optical theory and the mechanical theory is proposed. The aim is to avoid the complicated calculation process and provide a unique perspective approach for solving similar problems.
\section{Theoretical basis}
\begin{figure}[h!]
\centering
\includegraphics[width=0.8\textwidth]{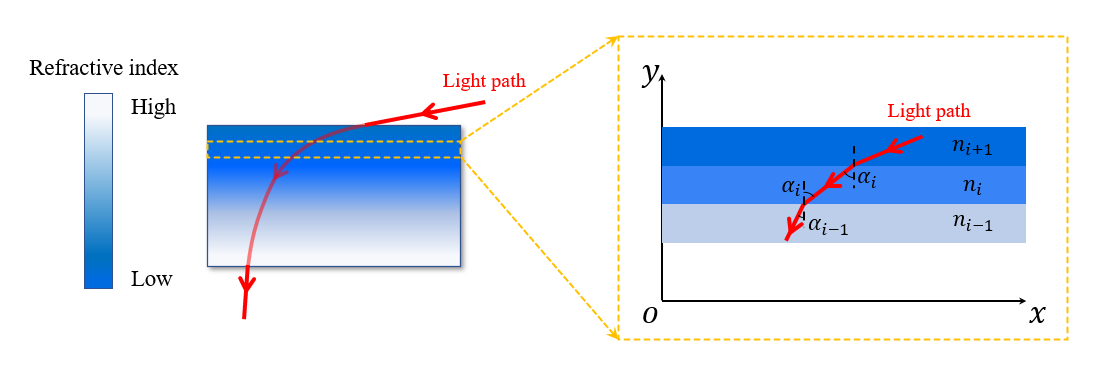}
\caption{Light path in the simplest continuously changing medium and the application of the infinitesimal method.}
\label{the simplest light path}
\end{figure}

In continuous inhomogeneous media, light rays propagate along curved paths. For the sake of simplicity, we assume that the refractive index varies only along the y-direction(Fig.\ref{the simplest light path}). Such a medium can be thought of as a limiting case of a medium consisting of a continuous set of thin slices of media of different refractive indices. At each interface, the light ray satisfies Snell’s law. In the spherically symmetrical medium model discussed in the following sections, this idea of discretization is applied.\cite{springer}

The other theoretical basis is the inverse-square central force system. The following kinematics and dynamics equations are crucial.
\begin{enumerate}
	\item The trajectory of a particle in the  inverse-square central force system is a conic curve, \cite{inverse} and the shape of the conic curve is determined by the energy of the particle $(E)$,\cite{basis} i.e.
	\begin{itemize}
		\item When $E=0, e=1$ , the orbit is a parabola.		
		\item When $E<0, e<1$ , the orbit is an ellipse.
		\item When $E>0, e>1$ , the orbit is a hyperbola.
		
	\end{itemize}
	\item The eccentricity of the conic curve is \cite{basis}
		\begin{equation}
			e=\sqrt{1+\frac{2h^2E}{mk^2}}     \label{e1}
		\end{equation}
\noindent
where $h=|\vec{r}|^2\dot{\theta}$,

$\vec{r}$: the position of the mass point relative to the center of force.

$\theta$: the angular displacement of the particle relative to the initial position.

$E$, $m$: the energy and the mass of the particle respectively.

$k$: the proportional coefficient of the inverse-square central force $F(r)=-\frac{km}{r^2}$.

	\item The equation of the conic section is \cite{basis}
		\begin{equation}
			r=\frac{\frac{h^2}{k}}{1+\sqrt{1+\frac{2h^2E}{mk^2}}cos(\theta-\theta_0)}     \label{e2}
		\end{equation}
\end{enumerate}
\section{Theoretical model}
\subsection{Geometry properties}

\begin{figure}[h!]
	\centering
	\includegraphics[width=0.9\textwidth]{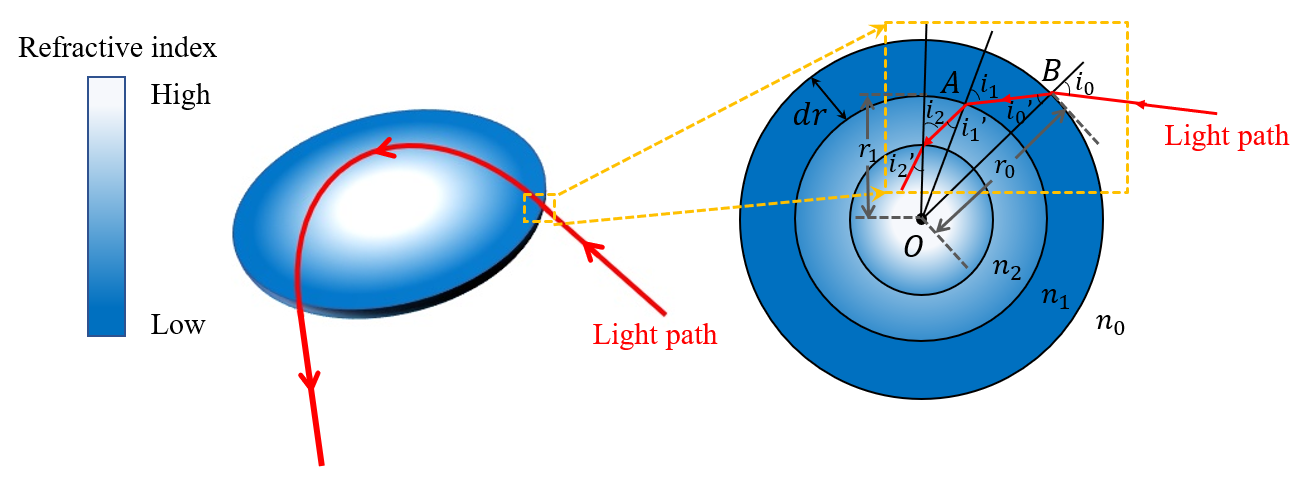}
	\caption{Light path in a layered structure of the medium model}
	\label{light path under infinitesimal}
\end{figure}

Based on the method of \textbf{section 2}, the medium model can be discretized into a continuous set of thin slices with different refractive indices(Fig.\ref{light path under infinitesimal}).

At each interface, the light ray satisfies Snell’s law and the following equation is obtained
\begin{equation}
	n_0sini_{0}=n_1sini_{0}',\quad n_1sini_{1}=n_2sini_{1}',\quad \ldots\label{e3-1}
\end{equation}

In $\bigtriangleup ABO$, by the law of sine
\begin{equation}
	\frac{sini_{0}'}{r_1}=\frac{sini_{1}}{r_0}	\label{e3-2}
\end{equation}

Substituting Eq.(\ref{e3-2}) into Eq.(\ref{e3-1}) yields
\begin{equation}
	r_0n_0sini_{0}=r_1n_1sini_{1}, \ldots	\label{e3-3}
\end{equation}

Similarly, we can state that the product
\begin{equation}
	r_0n_0sini_0=r_1n_1sini_1=r_2n_2sini_2=\ldots	\label{e3-4}
\end{equation}
is an invariant of the ray path; we will denote this invariant by $\beta$. It may be determined from the fact that if the ray initially makes an angle $i_0$ at a point where the refractive index is $n_0$, then the value of $\beta$ is $n_0sini_0$. Thus, in the limiting case of a continuous variation of the refractive index, the piecewise straight lines form a continuous curve(Fig.\ref{light path under infinitesimal}), which is determined from the equation
\begin{equation}
	r_in_isini=r_0n_0sini_0=\beta	\label{e3-5}
\end{equation}
implying that as the refractive index of this medium model changes, the ray path bends in such a way that the product $r_in_isini$ remains constant. Eq.(\ref{e3-5}) can be used to derive the ray equation. In fact, \textit{ chapter 14, Springer Handbook
of Lasers and Optics}\cite{springer} has already provided a detailed method of obtaining the PDEs of the medium with a continuous inhomogeneous refractive index, which shows a high level of mathematical and physical complexity for undergraduate students.

\subsection{Analogy relation equation}

\newcommand{\RNum}[1]{\uppercase\expandafter{\romannumeral #1\relax}}

Considering Eq.(\ref{e3-5}) and the law of the angular momentum conservation, we find that there is a high similarity in forms of mathematical equations. In fact, we can presume the propagation of light paths as the motion of a particle. The refractive index $n$ can be analogized to the momentum $p$ of the particle $m$. Similarly, the origin of the medium model should be the center of the inverse-square central force system. Therefore, the original optical problem can be transformed into a mechanical problem. The theory mentioned in \textbf{section 2} has proved that the motion trajectory of the particle $m$ is a conic curve in the inverse-square central force system. Hence, when the optical path presents a conic refraction in the medium with a spherically symmetric refractive index, the method of the inverse-square central force system can be used to study the relevant problems of the optical path quickly and concisely.

\begin{figure}[h!]
	\centering
	\includegraphics[width=0.6\textwidth]{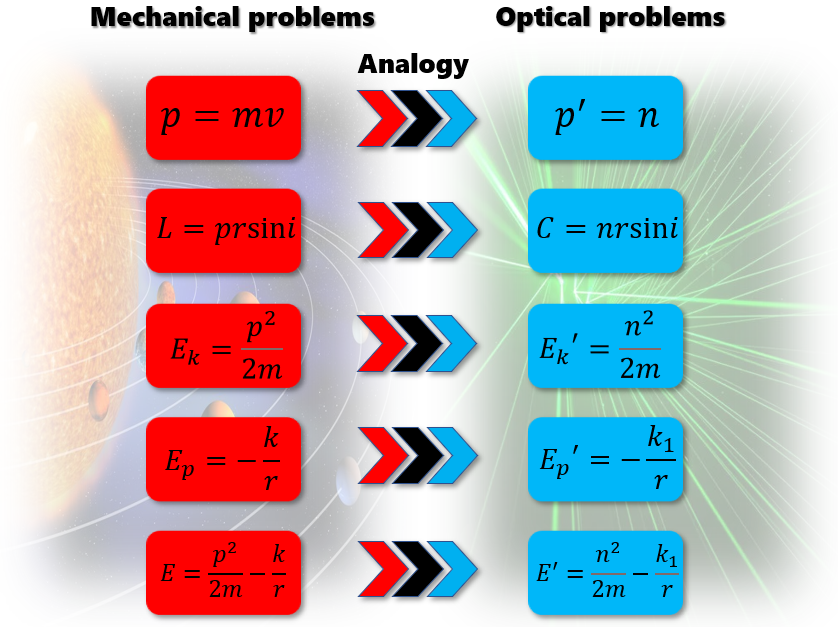}
	\caption{Analogue diagram of physical quantities.}
	\label{mindmap2}
\end{figure}

Fig.\ref{mindmap2} shows the analogy of the physical quantities in this optical and mechanical model.

\subsection{Qualitative analysis}
For ease of understanding, we define a virtual concept of ``feature point'' in our medium model, i.e. there is a feature point $J$ in the model if it satisfies the following three properties:
\begin{itemize}
    \item The ray is perpendicular to the $x$ axis of our coordinate system.
    \item The distance from the ray path to the origin of the coordinate is $|OJ|=r^*$.
    \item The refractive index of the feature point $J$ is $n^*$.
\end{itemize}
\begin{figure}[h!]
	\centering
	\includegraphics[width=0.6\textwidth]{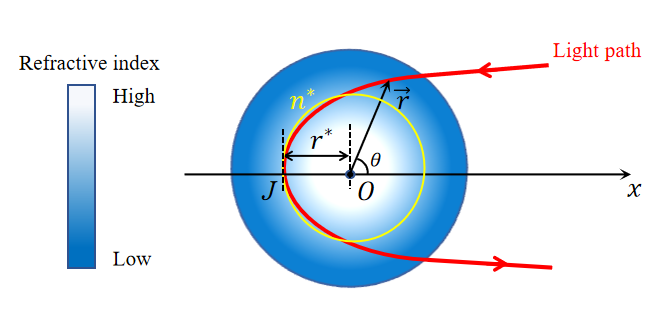}
	\caption{Light path diagram and the meaning of each physical quantity.}
	\label{light path with physical meaning}
\end{figure}

Based on \textbf{section 3.2}, the inverse-square central force system could be introduced since the light is refracted in a conical shape in the medium model. Thus we assume the origin $O$ as the center of the system.(Fig.\ref{mindmap2})

Considering the particle's kinetic energy $E_k'=\frac{n^2}{2m}$, potential energy $E_p'=-\frac{k_1}{r}$, we have the energy as a constant, $E'=\frac{n^2}{2m}-\frac{k_1}{r}$. We denote this invariant by $\lambda$, thus, $n^{2}=2 m \lambda+\frac{2 m k_{1}}{r}$. Obviously, the value of $\lambda$ can be determined by the feature point $J$ of the medium, i.e. $\lambda = \frac{n^{*2}}{2m}-\frac{k_1}{r^*}$. Then, the refractive index equation is
\begin{equation}
	n=n(r)=n^*\sqrt{1+A(\frac{1}{r}-\frac{1}{r^*})}	\label{e3-6}
\end{equation}
where $A=\frac{2mk_1}{n^{*2}}$.

According to the analogy of potential energy that in equation $E_p'=-\frac{k_1}{r}$, we have $k_1=GMm$. Therefore, the energy equation is
\begin{equation}
	E'=\frac{n^{*2}}{2m}-\frac{An^{*2}}{2mr^*}	\label{e3-7}
\end{equation}
i.e.
\begin{itemize}
	\item When $E^{'}=0$ , the orbit is a parabola.		
	\item When $E^{'}<0$ , the orbit is an ellipse.
	\item When $E^{'}>0$ , the orbit is a hyperbola.
\end{itemize}
\subsection{Quantitative analysis}
From Eq.(\ref{e3-7}), the value of $A$ determines the shape of the light path. Now we quantitatively discuss the relation between $A$ and the light path equation.

By setting the polar coordinate equation of the conic curve to be $r=\frac{\tilde{p}}{1-ecos\theta}$ and according to Eq.(\ref{e1}) and the principle of analogy, we have
\begin{align}
	e &=\sqrt{1+\frac{2E'L^2}{G^2M^2m^3}}\label{e3-8}\\
	  &=\sqrt{1+\frac{4r^*}{A^2}(r^*-A)}\label{e3-9}
\end{align}
where $L=n^*r^*$ (Fig.\ref{mindmap2}), $E'=\frac{n^{*2}}{2m}-\frac{An^{*2}}{2mr^*}$, $GM=\frac{An^{*2}}{2m}$ (Eq.(\ref{e3-7})).

Because $\tilde{p}=r^*(1-ecos\pi)=r^*(1+|\frac{2r^*}{A}-1|)$,
the curve equation is

\begin{equation}
	r=\frac{\tilde{p}}{1-ecos\theta}=\frac{(1+|\frac{2r^*}{A}-1|)r^*}{1-|\frac{2r^*}{A}-1|cos\theta}	\label{e3-10}
\end{equation}

Eq.~(\ref{e3-10}) shows the light path is a conic whose equation is determined by medium parameters $r^*$ and $A$. Eq.(\ref{e3-10}) demonstrates the results are consistent with \textbf{section 3.3}, since

\noindent
the orbit is a parabola when
\begin{equation}
    |\frac{2r^*}{A}-1|=1  \label{parabola}
\end{equation}
the orbit is an ellipse when
\begin{equation}
    |\frac{2r^*}{A}-1|<1  \label{ellipse}
\end{equation}
the orbit is a hyperbola when
\begin{equation}
    |\frac{2r^*}{A}-1|>1  \label{hyperbola}
\end{equation}

\section{Theoretical validation}
\subsection{The mathematical connection between mechanics and optics}
\begin{figure}[h!]
	\centering
	\includegraphics[width=0.6\textwidth]{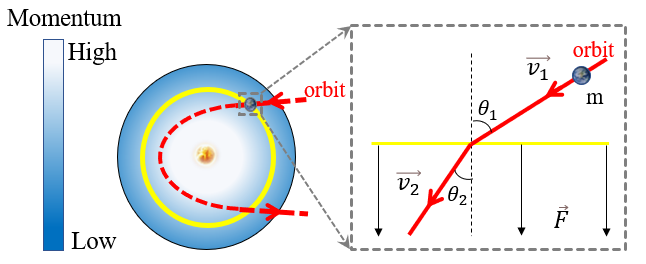}
	\caption{Particle in the inverse-square central force system.}
	\label{mechanical problem}
\end{figure}

We consider an equivalent mechanical problem of the inverse-square central force system (Fig.\ref{mechanical problem}), where a particle $m$ passes through an interface in the medium with velocity $\overrightarrow{v_1}$ and $\overrightarrow{v_2}$.

By the conservation of horizontal momentum ($p_x$) and the energy ($E$), we have
\begin{equation}
	mv_1sin\theta_1=mv_2sin\theta_2	\label{e4-1}
\end{equation}
\begin{equation}
	\frac{1}{2}mv_1^2+E_{p1}=\frac{1}{2}mv_2^2+E_{p2}=\gamma	\label{e4-2}
\end{equation}

\begin{figure}[h!]
	\centering
	\includegraphics[width=0.6\textwidth]{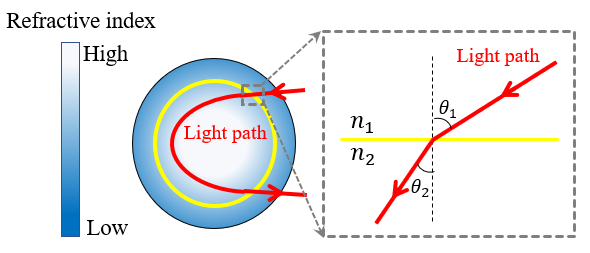}
	\caption{Light path in the medium with a concentric symmetric index.}
	\label{optical problem}
\end{figure}

Similarly, when it comes to an equivalent optical problem (Fig.\ref{optical problem}) where light passes through an interface with refractive index being $n_1$, $n_2$ respectively, we have Snell's Law $n_{1} \sin \theta_{1}=n_{2} \sin \theta_{2}$. Combined with Eq.(\ref{e4-1}) and Eq.(\ref{e4-2}), we could obtain Eq.(\ref{e4-3}) in the same form of Eq.(\ref{e4-2}), i.e.
\begin{equation}
	\frac{n_1^2}{2m}+\alpha^2E_{p1}=\frac{n_2^2}{2m}+\alpha^2E_{p2}=\alpha^2\gamma	\label{e4-3}
\end{equation}
where $\frac{n_{1}}{m v_{1}}=\frac{n_{2}}{m v_{2}}=\alpha$.

Eq.(\ref{e4-3}) shows that when using the analogy relationship of the physical quantities in Fig.\ref{mindmap2} to study optical problems, the physical meanings assigned are all based on strict mathematical derivations. The construction of analogy thinking is self-consistent in mathematics and physical logic.
\subsection{The theoretical nature of the optical-mechanical analogy}
Pierre de Fermat (1601-1665) postulated that 

\textit{the path between two points taken by a ray of light leaves the optical length stationary under variations in a family of nearby paths.}

Mathematically, Fermat's principle can be written as
\begin{equation}
  T=\int_{r_{1}}^{r_{2}} d t=\text { stationary } \label{fermat}  
\end{equation}
The points $r_{1}(x_1,y_1,z_1)$ and $r_{2}(x_2,y_2,z_2)$ are two fixed points in space. If we represent a general path $s$ from $r_1$ to $r_2$ in the form $x=x(z), y=y(z)$, with the velocity as $v$, and define the refractive index $n$ to be $n(x,y,z)=\frac{1}{v}$, Eq.(\ref{fermat}) becomes\cite{F_H}
\begin{equation}
    \int_{z_{1}}^{z_{2}} n(x, y, z) \sqrt{\dot{x}^{2}+\dot{y}^{2}+1} d z=\text { stationary } \label{fermat2}
\end{equation}

In 1744, Maupertuis proposed that ``Nature, in the production of its effects, does so always by the simplest means.''\cite{Maupertuis} Noether's theorem tells us that in a world without symmetry, the laws of physics change.\cite{Noether} In fact, the minimization of time revealed by Fermat's principle can remind us of the minimization of action stated by Hamilton’s principle, which is also based on the optical-mechanical analogy.\cite{hbook}

Hamilton's principle $(1834)$ or principle of least action can be written as:\cite{hbook2}

\textit{the motion of the system from fixed time $t_{1}$ to fixed time $t_{2}$ is such that the line integral (called the action or the action integral).}
\begin{equation}
    S=\int_{t_{1}}^{t_{2}} \mathcal{L}[x(t), \dot{x}(t), t] d t
\end{equation}
where $S$ is a functional of the function or path $x(t), \mathcal{L}$ is a Lagrangian system and $\dot{x}(t)=d x/d t$. $S$ has a stationary value for the actual path of the motion. $\delta S=0$ is the condition to have a minimum value.

For Eq.(\ref{fermat2}), in fact, we can identify the function
$$
\mathcal{L}(x, y, \dot{x}, \dot{y}, z)=n(x, y, z) \sqrt{\dot{x}^{2}+\dot{y}^{2}+1}
$$
as the equivalent of a Lagrangian system.\cite{F_H}
\begin{table}[h!]
\centering
\caption{Mathematical form of Fermat's principle and Hamilton’s principle}
\begin{ruledtabular}
\begin{tabular}{c c}
Fermat's principle & Hamilton’s principle \\
\hline	
$\int_{r_{1}}^{r_{2}} d t=\int_{r_{1}}^{r_{2}} n(x, y, z) d s=\text { stationary }$  & $\int_{t_{1}}^{t_{2}} \mathcal{L}[x(t), y(t), \dot{x}(t), \dot{y}(t), t] d t=\text { stationary }$ \\
$\int_{z_{1}}^{z_{2}} \mathcal{L}[x(z), y(z), \dot{x}(z), \dot{y}(z), z] d z=\text { stationary }$ & {}  \\
\end{tabular}
\end{ruledtabular}
\label{t1}
\end{table}

We see from Table.\ref{t1} that Fermat's principle takes a stationary value for a function of a length coordinate and Hamilton's principle takes a stationary value for functions of time. Or, Fermat's principle takes a minimum value for the transit time with the endpoints, $r_{1}, r_{2}$, fixed in space, whereas Hamilton's principle takes a minimum value for the action with the endpoints, $\left(r_{l}, t_{l}\right),\left(r_{2}, t_{2}\right)$ fixed in both space and time.\cite{F_H}

Therefore, Fermat's principle is equivalent to Hamilton's principle mathematically, which essentially reveals the correctness of the analogy method.

Moreover, this paper \cite{analogy} also demonstrates the two different interpretations of Hamilton’s characteristic function in optics and mechanics, which could also further verify the underlying principles.
\section{Computer Simulation}
The description of the laws of physics for space- and time-dependent problems are usually expressed in terms of partial differential equations (PDEs), which cannot be solved with analytical methods. Instead, an approximation of the equations can be constructed, typically based upon different types of discretizations, which can be computed numerically through the finite element method (FEM). And this can also be applied to the curved ray equation in the aforementioned medium model.

COMSOL Multiphysics is based on the FEM. It solves PDEs to achieve the simulation of real physical phenomena and uses mathematical methods to solve the physical phenomena in the real world. The propagation of light is modelled in the geometrical optics approximation. Thus, the method of choice is ray tracing. Non-sequential tracing is used, in which no predefined paths are assumed for rays. COMSOL Multiphysics Ray Tracing module is based on Hamiltonian optics and the FEM. The ray tracer is time-dependent and employs a Monte-Carlo approach.\cite{comsol}

Here we modeled the concentric symmetric continuous inhomogeneous medium and established the rectangular coordinate system with the origin as the center of the medium. Now the goal is to verify Eq.(\ref{parabola}) - Eq.(\ref{hyperbola}) obtained through analogy method, which defines the relation between conic shapes and medium parameters. The input-output parameters are shown as Table \ref{t2}.
\begin{table}[h!]
\centering
\caption{The input-output parameters in COMSOL Multiphysics}
\begin{ruledtabular}
\begin{tabular}{c c c}
\textbf{Input} & \textbf{Intermediate variable} & \textbf{Output} \\
\hline	
$n(r), n(x,y)$ & {} & {}   \\
$r_0$ & {} & {} \\
$n_0$ & {$n^*$} & {light path and its coordinates}  \\
$T_s$ & {} & {}  \\
Exit point of light $(x_0,y_0)$ & {} & {} \\
$\{(r_i^*,A_i)|f(r_i^*,A_i)=0,i=1,2,3,\ldots\}$  & {} & {}  
\end{tabular}
\end{ruledtabular}
\label{t2}
\begin{tablenotes}
\footnotesize
    \item[$^1$] $^1$$n(r) / n(x,y)$(polar/rectangular coordinate system) is the equation of the refractive index distribution. 
    \item[$^2$] $^2$$r_0$ is the maximum radius of the concentric symmetric medium. 
    \item[$^3$] $^3$$n_0$ is the refractive index of the periphery of the medium.
    \item[$^4$] $^4$$T_s$ is the simulation time. 
    \item[$^5$] $^5$$(x_0,y_0)$: For the sake of simplicity, we set the feature point $J(-r^*,0)$ of the medium as the exit point of the light ray.  
    \item[$^6$] $^6$$\{(r_i^*,A_i)|i=1,2,3,\ldots\}$ is series of different parameters combinations.
    \item[$^7$] $^7$$n^*$ is the refractive index value at the feature point $J$.
\end{tablenotes}
\end{table}

\begin{figure}[htbp]
\centering
\subfigure[Parabolic light path]
{
	\begin{minipage}{4cm}
	\centering
	\includegraphics[scale=0.16]{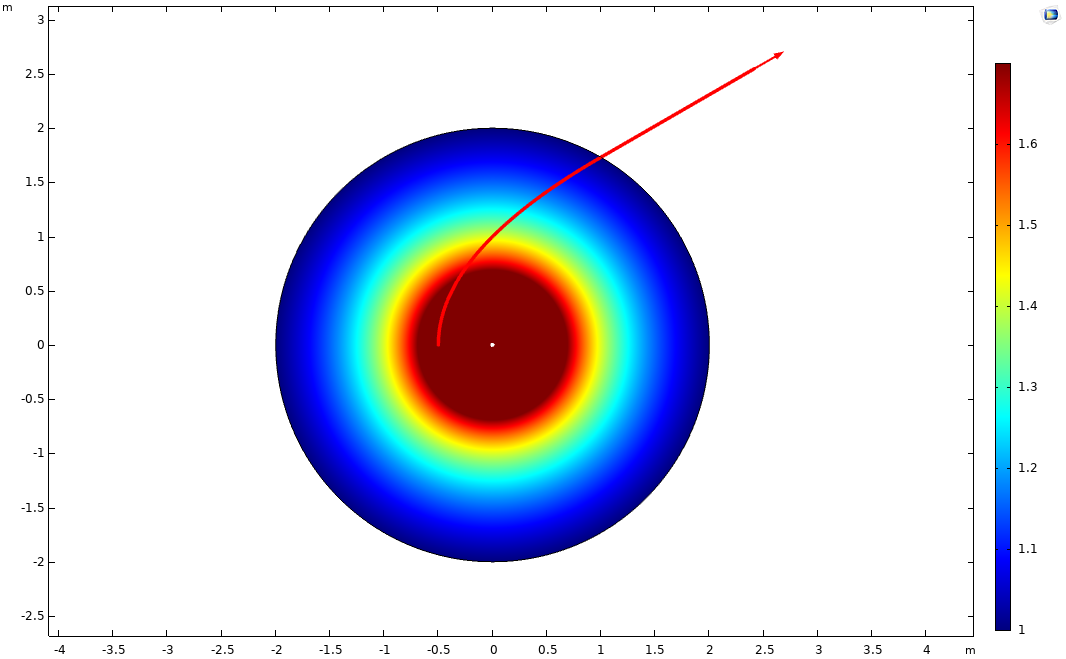}
	\end{minipage}
}
\subfigure[Elliptical light path]
{
	\begin{minipage}{4cm}
	\centering
	\includegraphics[scale=0.16]{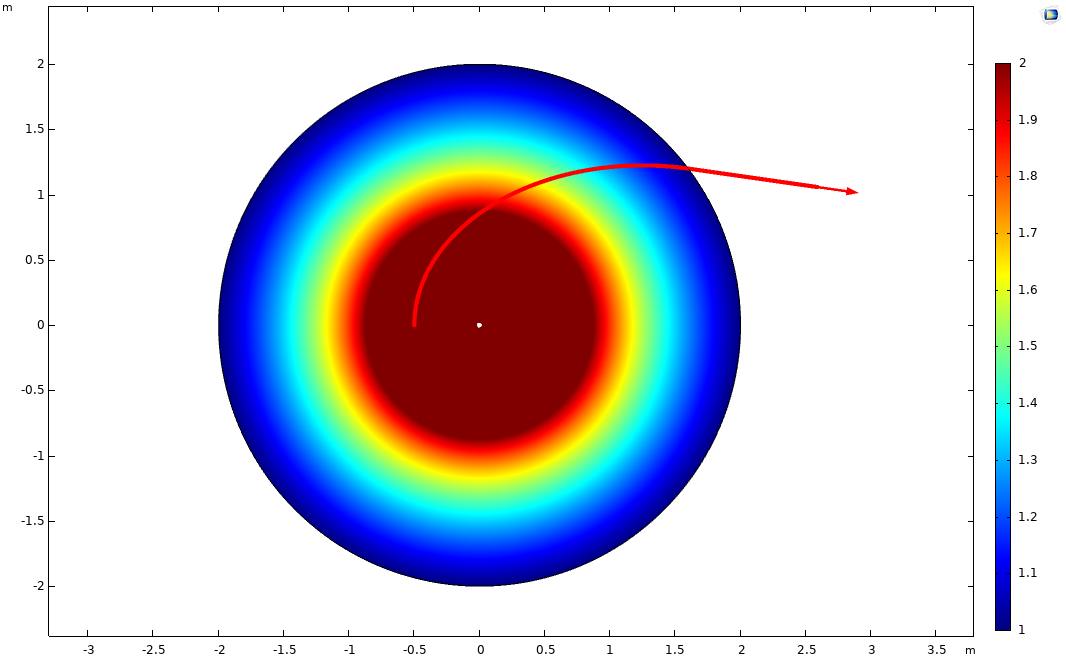}
	\end{minipage}
}
\subfigure[Hyperbolic light path]
{
	\begin{minipage}{4cm}
	\centering
	\includegraphics[scale=0.16]{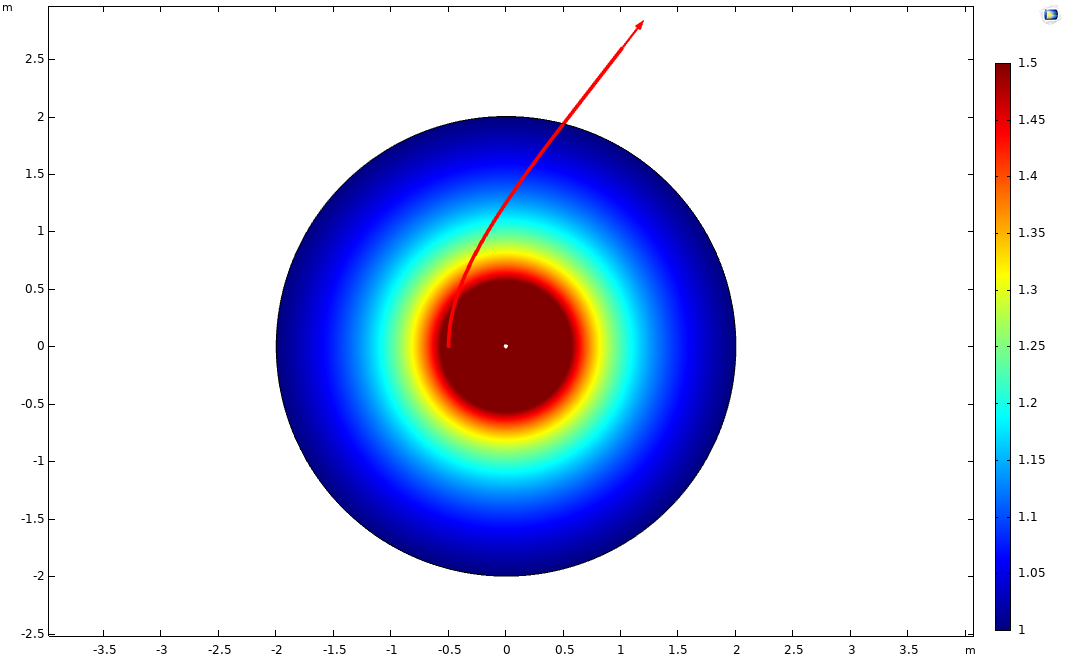}
	\end{minipage}
}
\caption{Comparison between simulation results and theory.}
\label{sim}
\end{figure}

From Eq.(\ref{parabola}) - Eq.(\ref{hyperbola}), the following conclusions hold:
\begin{itemize}
    \item The light path is a parabola if: $|\frac{2r^*}{A}-1|=1 \Longrightarrow r^*=A, (r^*<r_0)$
    \item The light path is an ellipse if: $|\frac{2r^*}{A}-1|<1 \Longrightarrow 0<r^*<A, (r^*<r_0)$
    \item The light path is a hyperbola if: $|\frac{2r^*}{A}-1|>1 \Longrightarrow r^*>A, (r^*<r_0)$ 
\end{itemize}
In fact, we can take any value of the input that satisfies the corresponding restrictions. Here we define a set of the input $\{\text{'parabola'},\text{'ellipse'},\text{'hyperbola'}\}$ as
\begin{align*}
    &r_0 = \{2,2,2\} (meters)\\
    &n_0 = \{1,1,1\} (meters)\\
    &T_s = \{17,20,12\} (ns)\\
    &\text{Exit point} (x,y): \{(-0.5,0),(-0.5,0),(-0.5,0)\} (meters)\\
    &\{(r^*,A)\vert|f(r^*,A)=0\}=\{(0.5,0.5),(0.5,\frac{7}{12}),(0.5,\frac{2}{5})\}\\
    &
    \begin{cases}
    n(r)_{\text{parabola}} = n^*\sqrt{1+A(\frac{1}{r}-\frac{1}{r^*})}=\sqrt{\frac{2}{r}} \Longrightarrow n(x,y)=\sqrt{2}(x^2+y^2)^{-\frac{1}{4}}\\
    n(r)_{\text{ellipse}} = n^*\sqrt{1+A(\frac{1}{r}-\frac{1}{r^*})}=\sqrt{\frac{14}{3}\cdot\frac{1}{r}-\frac{4}{3}} \Longrightarrow n(x,y)=\sqrt{\frac{14}{3\sqrt{x^2+y^2}}-\frac{4}{3}}\\
    n(r)_{\text{hyperbola}} = n^*\sqrt{1+A(\frac{1}{r}-\frac{1}{r^*})}=\sqrt{\frac{1}{r}+\frac{1}{2}} \Longrightarrow n(x,y)=\sqrt{\frac{1}{\sqrt{x^2+y^2}}+\frac{1}{2}}
    \end{cases}
\end{align*}
where
\begin{align*}
    \begin{cases}
    (n_0=1,r_0=2)_{\text{parabola}} \in \{(n,r)\vert n - n^*\sqrt{1+A(\frac{1}{r}-\frac{1}{r^*})}=0, r^*=A\} 
    \Longrightarrow n^*=2\\
    (n_0=1,r_0=2)_{\text{ellipse}} \in \{(n,r)\vert n - n^*\sqrt{1+A(\frac{1}{r}-\frac{1}{r^*})}=0, 0<r^*<A\} 
    \Longrightarrow n^*=2\sqrt{2}\\
    (n_0=1,r_0=2)_{\text{hyperbola}} \in \{(n,r)\vert n - n^*\sqrt{1+A(\frac{1}{r}-\frac{1}{r^*})}=0, A<r^*<r_0\} 
    \Longrightarrow n^*=\frac{\sqrt{10}}{2}
    \end{cases}
\end{align*}
Filting out the secondary reflections, the outputs (the light path) are shown in Fig.\ref{sim}.

\begin{figure}[htbp]
\centering
\subfigure[]
{
	\begin{minipage}{4.5cm}
	\centering
	\includegraphics[scale=0.25]{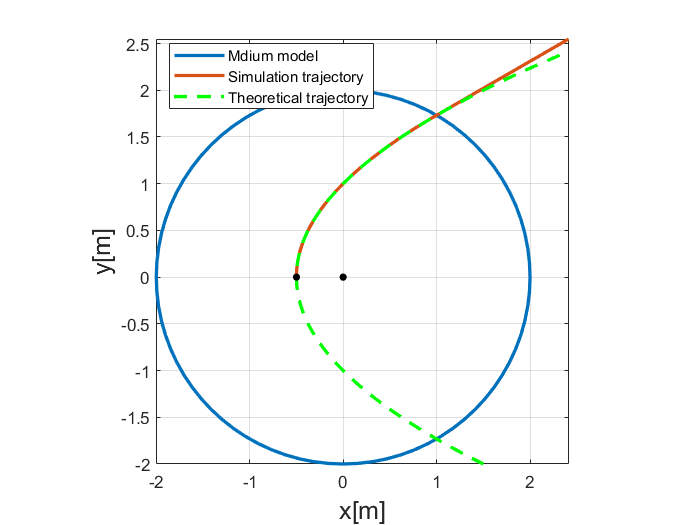}
	\end{minipage}
}
\subfigure[]
{
	\begin{minipage}{4.5cm}
	\centering
	\includegraphics[scale=0.25]{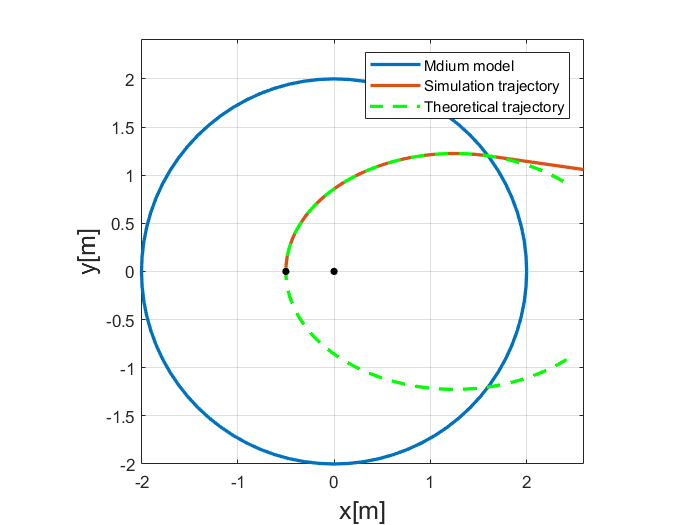}
	\end{minipage}
}
\subfigure[]
{
	\begin{minipage}{4.5cm}
	\centering
	\includegraphics[scale=0.25]{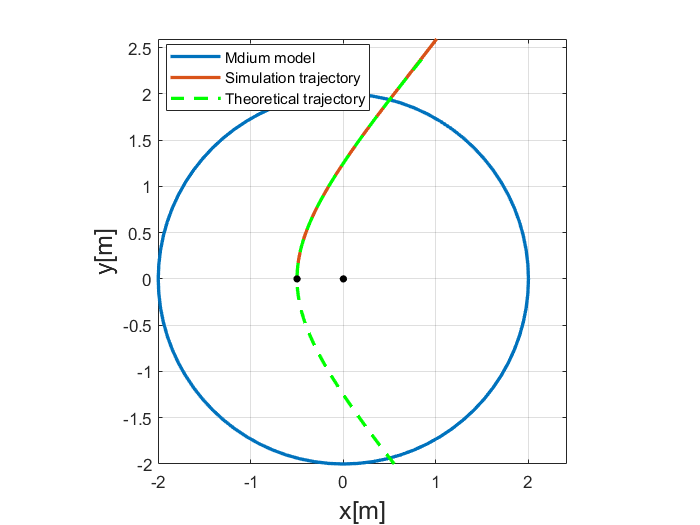}
	\end{minipage}
}
\subfigure[Parabola]
{
	\begin{minipage}{4.5cm}
	\centering
	\includegraphics[scale=0.25]{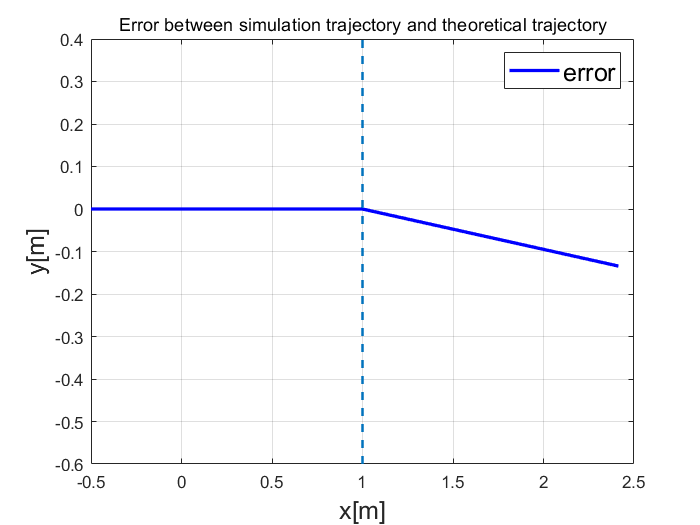}
	\end{minipage}
}
\subfigure[Ellpise]
{
	\begin{minipage}{4.5cm}
	\centering
	\includegraphics[scale=0.25]{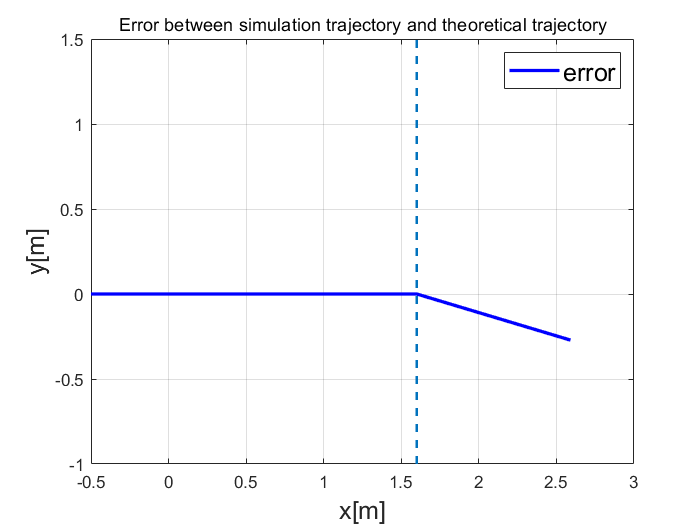}
	\end{minipage}
}
\subfigure[Hyperbola]
{
	\begin{minipage}{4.5cm}
	\centering
	\includegraphics[scale=0.25]{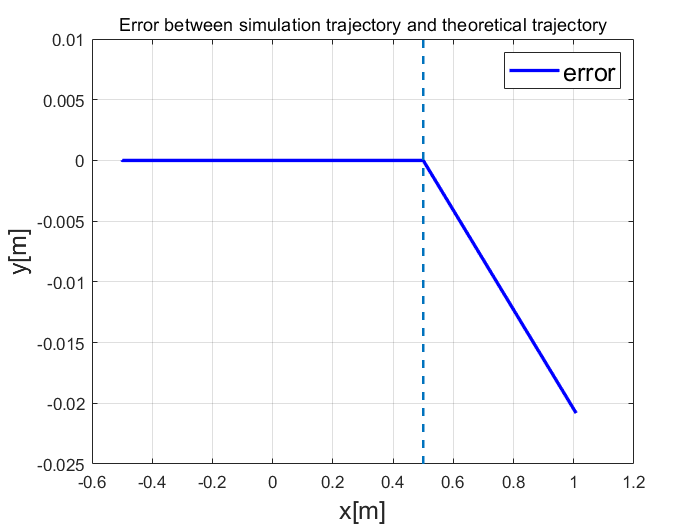}
	\end{minipage}
}
\caption{Trajectory contrast (a)(b)(c) and their corresponding error graphs (d)(e)(f) between simulation and theoretical results of three cases.}
\label{plot}
\end{figure}

Now we analyse the quantitative outputs (coordinates of simulation results) and perform data visualization on it. The theoretical trajectory equation can be obtained from Eq.(\ref{e3-10}), i.e.
$$
    \text{parabola: } r=\frac{1}{1-cos\theta} \Longrightarrow \sqrt{x^2+y^2}-x-1=0
$$
$$
    \text{ellipse: } r=\frac{\frac{6}{7}}{1-\frac{5}{7}cos\theta} \Longrightarrow \sqrt{x^2+y^2}-\frac{5}{7}x-\frac{6}{7}=0 
$$
$$
    \text{hyperbola: } r=\frac{5}{4-6cos\theta} \Longrightarrow 2\sqrt{x^2+y^2}-6x-5=0
$$
Thus, Fig.\ref{plot} shows that the error is $0$ when the light ray is inside the medium model, i.e., the simulation results of three different light paths are consistent with the theory.
\section{Conclusion}
Consider that in daily life, the problem of the curve propagation of the light path such as the laser ranging, photoelectric detection, astronomical observation and other fields referred to in the introduction are often encountered. In fact, in the earth's atmosphere, spherical solutions are possible for certain ideal conditions with specific lapse rates. This proves the possibly of real situations that fulfill the requirements for the conic path trajectories as discussed. Morever, based on the equations in this paper, artificial media causing a conic light path can also be easily designed theoretically. In all, when the optical path is needed to be made a conic curve, the inverse-square central force system can be used for analogy study. When the optical path is needed to be made a special curve, the inverse-cubic or even the $n^{th}$ inverse central force system can also be used to solve the problem similarly.Thus we can extend the application of such methods to more general applications, which is an important application example of the mechanical analogy of optical principles.



\begin{thebibliography}{99}
\bibitem{feynman}Robert B Leighton and Matthew Sands. The Feynman lectures on physics. Addison-Wesley Boston, MA, USA, 1965.
\bibitem{laser}Haojian Yan and Guangli Wang. New consideration of atmospheric refraction in laser ranging data. Monthly Notices of the Royal Astronomical Society, 307(3):605–610, 1999.
\bibitem{photoelectric} Benjamin K Stuhl. Atmospheric refraction corrections in ground-to-satellite optical time transfer. Optics Express, 29(9):13706–13714, 2021.
\bibitem{astronomical} Andrew Tipton Young. Understanding astronomical refraction. The Observatory, 126:82–115, 2006
\bibitem{hecht} Eugene Hecht. Optics. Pearson Education India, 2012.
\bibitem{lehn} Waldemar H Lehn and W Friesen. Simulation of mirages. Applied optics, 31(9):1267–1273, 1992.
\bibitem{lehn1985} Waldemar H Lehn. A simple parabolic model for the optics of the atmospheric surface layer. Applied mathematical modelling, 9(6):447–453, 1985.
\bibitem{atmosphere} Charles-Antoine L’Hour, Vincent Fabbro, Alexandre Chabory, and Jerome Sokoloff. 2-d propagation modeling in inhomogeneous refractive atmosphere based on gaussian beams part ii: Application to radio occultation. IEEE TRANSACTIONS ON ANTENNAS AND PROPAGATION, 67(8, 2):5487–5496, AUG 2019.
\bibitem{white} Hua-ming Qian, Long Sun, Jia-nan Cai, and Wei Huang. A starlight refraction scheme with single star sensor used in autonomous satellite navigation system. Acta Astronautica, 96:45–52, 2014.
\bibitem{qian} Robert L White, Sam W Thurman, and Frank A Barnes. Autonomous satellite navigation using observations of starlight atmospheric refraction. Navigation, 32(4):317–333, 1985.
\bibitem{wu} Yanxiong Wu, Xin Zhang, Jizhen Zhang, Lingjie Wang, Hemeng Qu, Yang Zhu, and Fei Zeng. Research on the autonomous star sensor based on indirectly sensing horizon and its optical design. Acta Optica Sinica, 35(2):0222001.
\bibitem{ricklin} Jennifer C Ricklin, Stephen M Hammel, Frank D Eaton, and Svetlana L Lachinova. Atmospheric channel effects on free-space laser communication. Journal of Optical and Fiber Communications Reports, 3(2):111–158, 2006.
\bibitem{liu} H Liu, L Lu, BF Zhang, CX Wu, and JY Wang. Effect of atmospheric refraction on timing deviation of earth-to-satellite time transfer based on dual wavelength. Frontier Research and Innovation in Optoelectronics Technology and Industry, K. Habib and E. Lewis, eds.(Taylor \& Francis Group, 2019) pp, pages 319–325, 2018.
\bibitem{math} Frank Trager. Springer handbook of lasers and optics,  volume 2. Springer, 2012.
\bibitem{springer} Chuangjie Xu, Jinhong Wu, You Wu, Ludong Lin, Jianbin Zhang, and Dongmei Deng. Propagation of the pearcey gaussian beams in a medium with a parabolic refractive index. Optics Communications, 464:125478, 2020.
\bibitem{inverse} Don Chakerian. Central force laws, hodographs, and polar reciprocals. Mathematics Magazine, 74(1):3–18, 2021.
\bibitem{basis} Pini Gurfil and P Kenneth Seidelmann. Central force motion. In Celestial Mechanics and Astrodynamics: Theory and Practice, pages 79–94. Springer, 2016.
\bibitem{F_H} Malcolm Anderson, Miftachul Hadi, and UA Deta. Fermat’s principle and hamilton’s principle: Does a least action take a least time for happening? In Journal of Physics: Conference Series, volume 1467, page 012038. IOP Publishing, 2020.
\bibitem{Maupertuis}Alberto G. Rojo. Hamilton’s principle: Why is the integrated difference of the kinetic and potential energy minimized? American Journal of Physics, 73(9):831–836, 2005.
\bibitem{Noether} Iman Marvian and Robert W. Spekkens. Extending noether’s theorem by quantifying the asymmetry of quantum states. NATURE COMMUNICATIONS, 5, MAY 2014.
\bibitem{hbook} William Rowan Hamilton. On a General Method of Expressing the Paths of Light, \& of the Planets, by the Coefficients of a Characteristic Function. PD Hardy, 1833.
\bibitem{hbook2} H Goldstein, CP Poole, and J Safko. edition 3. classical mechanics, 2000.
\bibitem{analogy} Christian Joas and Christoph Lehner. The classical roots of wave mechanics: Schrodinger’s transformations of the optical-mechanical analogy. Studies in History and Philosophy of Science Part B: Studies in History and Philosophy of Modern Physics, 40(4):338–351, 2009.
\bibitem{comsol} Mihails Birjukovs, Aleksandrs Jegorovs, Andris Jakovics, Aleksejs Klokovs, and Ingmars Felcis. Design optimization automation for luminaire reflectors using comsol multiphysics and performance comparison against zemax opticstudio. In Nikitov, SA and Bykov, DE and Borovik, SY and Pleshivtseva, YE, editor, 2019 XXI INTERNATIONAL CONFERENCE COMPLEX SYSTEMS: CONTROL AND MODELING PROBLEMS (CSCMP), pages 208–213. Russian Acad Sci, Inst Control Complex Syst; Samara State Tech Univ, Volga Reg, 2019. 21st International Conference on Complex Systems - Control and Modeling Problems (CSCMP), Samara, RUSSIA, SEP 03-06, 2019.

\end{thebibliography}

\end{document}